\documentclass[12pt]{article}
\usepackage{graphicx}
\usepackage{amsmath}
\usepackage{amssymb}
\usepackage{bbm}

\usepackage{color}
\usepackage{psfrag}

\newcommand{\be}{\begin{equation}}
\newcommand{\ee}{\end{equation}}
\newcommand{\bes}{\begin{equation*}}
\newcommand{\ees}{\end{equation*}}

\newcommand{\Dslash}{\mbox{$D$\kern-0.65em \hbox{/}\hspace*{0.25em}}}
\newcommand{\Dslashup}{\not \kern-0.03em D}
\def\slashZ#1{#1\!\!\!\! \boldmath{\times} \!\!\!\,\,}
\def\slashN#1{#1\!\!\!\!\!\! \boldmath{\times}  \!\,\,}

\begin{document}

\begin{titlepage}

\begin{center}

\begin{flushright}
 CERN-PH-TH-2006-007\\
 hep-lat/0602005\\
\end{flushright}

\textbf{\large QCD at Zero Baryon Density \\ and the Polyakov Loop Paradox \\[6ex]}

{Slavo Kratochvila$^{a,}$\footnote{skratoch@itp.phys.ethz.ch}
 and Philippe de Forcrand$^{ab,}$\footnote{forcrand@itp.phys.ethz.ch}}
\\[4ex]
{${}^a${\it Institute for Theoretical Physics, ETH Z\"{u}rich,
CH-8093 Z\"{u}rich, Switzerland}\\[1ex]
${}^b${\it CERN, Physics Department, TH Unit, CH-1211 Geneva 23, Switzerland}}
\\[8ex]
{\small \bf Abstract}\\[2ex]
\begin{minipage}{14cm}
{\small

We compare the grand canonical partition function 
at fixed chemical potential $\mu$ with the
canonical partition function 
at fixed baryon number $B$, formally and by numerical simulations
at $\mu=0$ and $B=0$ with four flavours of staggered quarks. We verify
that the free energy densities are equal in the thermodynamic limit, and show that they can be well
described by the hadron resonance gas at $T < T_c$ and by the free
fermion gas at $T>T_c$.\\

Small differences between the two ensembles, for thermodynamic observables
characterising the deconfinement phase transition, vanish with increasing lattice size.
These differences are solely caused by contributions of non-zero baryon density
sectors, which are exponentially suppressed with increasing volume. The Polyakov
loop shows a different behaviour: for \emph{all} temperatures and volumes, 
its expectation value is exactly zero in the canonical formulation, whereas it is always
non-zero in the commonly used grand-canonical formulation. We clarify this  paradoxical
difference, and show that
the non-vanishing Polyakov loop expectation value is due to contributions
of non-zero triality states, which are not physical, because they give zero 
contribution to the partition function.

}

\end{minipage}
\end{center}
\vspace{1cm}

\end{titlepage}

\section{Introduction}
\label{sec:introduction}

To simulate QCD thermodynamics on the lattice~\cite{Fodor:2001au}, 
one commonly uses the grand canonical partition function
with respect to the quark number as a function of a chemical potential $\mu$. It is given, after integration over
the fermion fields, by
\be \label{eq:introduction_ZGC}
Z_{ GC }(T,\mu) = \int [DU] \; e^{-S_g[\beta,U]} \det M(U;\mu)\;.
\ee
Recently, another approach using a canonical formalism has been used~\cite{Kratochvila:2004wz,Kratochvila:2005mk}.
The canonical partition function, which will be derived in detail in the next section, is
\be \label{eq:introduction_ZC}
Z_{ C}(T,Q) = \int_{-\infty}^{\infty} d \left( \frac{ \mu_I }{ T }\right)\;
e^{-i Q \frac{ \mu_I }{ T }} Z_{ GC }(T,\mu =  i \mu_I )\;.
\ee
The physics described by both ensembles, grand-canonical and canonical, must be
the same in the thermodynamic limit, ie.~the free energy density should be the same. This has been
shown in Ref.~\cite{Azcoiti:1998rj,Kratochvila:2003rp}, and will be confirmed in this study.
It is thus puzzling that the Polyakov loop expectation value in the canonical ensemble is
exactly zero, while it is non-zero in the grand-canonical ensemble, for all temperatures and volumes.
We will show that this discrepancy is due to contributions from the canonical sectors with a quark
number that is not a multiple of three: the so-called non-zero triality sectors\footnote{\emph{Triality}
is defined as the difference between the number of quarks and the number of anti-quarks modulo 3.}. 
Discussions on the role of these non-zero triality sectors have a long history~\cite{Oleszczuk:yg}
and include speculations that their influence persists even in the thermodynamic limit, so
that they must be explicitly projected out. It is our purpose to clarify these issues.

In the following, we discuss properties of the grand canonical partition function as a function
of an imaginary chemical potential and construct the canonical partition function (Section II). 
We then show that the Polyakov loop vanishes in the canonical ensemble (Section III), and
resolve the paradox above (Section IV). 
After presenting our numerical method to simulate the zero baryon density sector, which is the first step 
towards finite density QCD simulations (Section V), we elaborate on the results for the free
energy density as a function of the imaginary chemical potential, which we compare with predictions of the
hadron resonance gas model~\cite{Hagedorn:1980cv} and of a free fermion gas (Section VI). 
We further study the expectation values and finite size effects of thermodynamic observables, 
like the plaquette and the chiral condensate in both formulations (Section VII). Conclusions follow.
Preliminary results of this study have been presented in Ref.~\cite{Kratochvila:2003rp}.

\section{Canonical Ensemble}
\label{sec:canonicalensemble}

\begin{figure}[!htb]
\begin{center}
    \includegraphics[angle=-90,width=12.0cm]{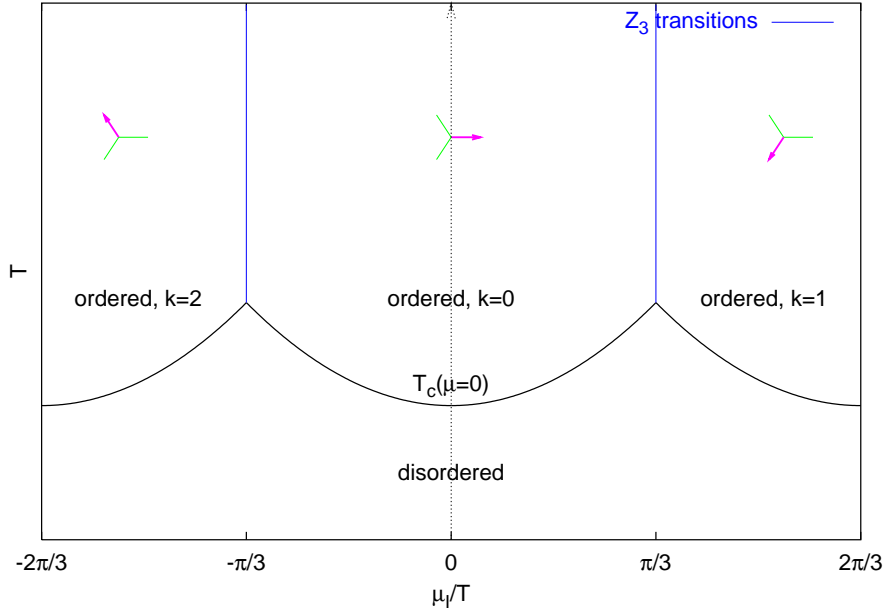}
  \caption{Phase diagram of $Z_{GC}(i\mu_I)$ in the $(\mu_I,T)$ plane. The arrows indicate the orientation
  of the Polyakov loop.
  The vertical lines mark the ``order-order'' $Z_3$ transitions, which are first-order.
  Properties of the ``order-disorder'' $Z_3$ transitions (curved lines) depend on the parameters
  (number of flavours, quark masses) of the theory.}
  \label{fig:ensemble_phasediagram}
\end{center}
\end{figure}

Let us first discuss symmetries of the grand canonical partition function  $Z_{GC}(T,\mu)$ as a function of
an imaginary chemical potential $\mu= i \mu_I$, following Ref.~\cite{Roberge:mm}:
\begin{itemize}
\item Evenness: $Z_{GC}(-i \mu_I) = Z_{GC}(+i \mu_I)$.\\
The transformation $\mu \to - \mu$ can be compensated by time-reversal, ie.~by interchanging particles and anti-particles.
Time reversal is equivalent to CP symmetry (since CPT is always a good symmetry), and thus does not change the thermodynamics in the absence of CP violating terms.
\item $\frac{2\pi T}{3}$-periodicity in $\mu_I$: ${Z_{GC}(i( \mu_I + \frac{2\pi T}{3}))} = Z_{GC}(i \mu_I)$.\\
A shift in the imaginary chemical potential $\mu_I \to \mu_I + \frac{2 \pi k}{3} T$ can be exactly compensated by
a $Z_{3}$-transformation, ie.~by a rotation of the Polyakov loop $Pol({\vec x}) \to e^{-i\frac{2\pi k}{3}} Pol({\vec x})$,
configuration by configuration.
Since the path integral sums over all possible gauge fields, the partition function stays the same.
\end{itemize}
These properties lead to an expectation for the phase structure in the $T-\mu_I$ plane, see 
Ref.~\cite{Roberge:mm} and Fig.~\ref{fig:ensemble_phasediagram}.
In the low-temperature phase where the $Z_{3}$-symmetry is realised (``disordered''), the periodicity 
in $\mu_I$ is smoothly realised. 
However, at high temperature,
this $Z_{3}$-symmetry is spontaneously broken in favor of one of the $Z_3$-sectors (``ordered''). 
Which sector is preferred depends on $\mu_I$. Thus, the existence of $Z_3$-sectors
forces the appearance of phase transitions at high temperature. 
By symmetry, these discontinuities must appear at $\mu_I=\frac{2 \pi k T}{3}+\frac{\pi T}{3}$.
The transitions are first-order since the derivative of the free energy with respect to $\mu_I$
is discontinuous. 
\\

To obtain the canonical partition function $Z_C(T,Q)$, we fix to $Q$ the conserved charge 
$\hat N=\int d^3 \vec x \; \bar{\psi}(\vec x) \; \gamma_0 \; \psi(\vec x)$,
which represents the net quark number. This is accomplished 
by inserting a $\delta$-function in the grand canonical partition function:
\be
Z_{C}(T,Q) = \int [DU][D\bar\Psi][D\Psi]\; e^{-S_g[U;T]-S_F[U,\bar \Psi, \Psi;T]}  \delta\left(\hat N-Q\right)\;.
\ee
The $\delta $-function admits a Fourier representation with the new variable $\bar \mu_I$:
\begin{align}
Z_{C}(T,Q) & = \mathcal{N} \int_{-\infty}^{\infty} d\bar \mu_I  e^{-i Q \bar \mu_I} \times \int [DU][D\bar\Psi][D\Psi]\;
        e^{-S_g[U;\beta]-S_F[U,\bar{\psi},\psi] + i  \bar \mu_I \hat N } \notag \\
           & = \mathcal{N} \int_{-\infty}^{\infty} d\bar \mu_I  e^{-i Q \bar \mu_I} \times \int [DU][D\bar\Psi][D\Psi]\; \times \notag  \\
           & \qquad \qquad\times e^{-S_g[U;\beta]-S_F[U,\bar{\psi},\psi] + i  \bar \mu_I \int d^3 \vec x \; \bar{\psi}(\vec x) \; \gamma_0 \; \psi(\vec x) } \notag  \\
          & = \mathcal{N} \int_{-\infty}^{\infty} d\bar \mu_I  e^{-i Q \bar \mu_I} \times \int [DU][D\bar\Psi][D\Psi]\; \times \notag  \\
           &  \qquad\qquad \times e^{-S_g[U;\beta]-S_F[U,\bar{\psi},\psi] + i  \bar \mu_I T \int_0^{\frac{1}{T}} d\tau \int d^3 \vec x \; \bar{\psi}(\vec x) \; \gamma_0 \; \psi(\vec x) } \;.
\end{align}
where $\mathcal{N}$ is a normalisation factor. In the last line, we have used the fact that $Q$ is conserved.

One recognises $i \mu_I = i \bar \mu_I T$  as an imaginary chemical potential, so that
\be
  Z_{C}(T,Q) =\mathcal{N} \int_{-\infty}^{\infty} d\bar \mu_I  e^{-i Q \bar \mu_I } Z_{GC}(T,i\bar \mu_I T  )
  = \frac{3}{2\pi} \int_{-\frac{\pi}{3}}^{\frac{\pi}{3}} d\bar \mu_I  e^{-i Q \bar \mu_I }Z_{GC}(T,i\bar \mu_I T  )\;,
\label{eq:ZC_Q}
\ee
where we have exploited the  $\frac{2\pi T}{3}$-periodicity in $\mu_I$ of $Z_{GC}(i\mu_I )$ in the last
step\footnote{Note that the evenness of $Z_{GC}(i\mu_I )$ in  $\mu_I$ implies $Z_{C}(T,Q)=Z^*_{C}(T,-Q)$.
In particular the $Z_C(T,Q)$'s are real as expected.}. From this periodicity,
it follows that \emph{$Z_{C}(T,Q)=0$ except for $\frac{Q}{3} \equiv B \in {\mathbb Z}$}, where $B$ is the baryon number.
The canonical partition function \emph{vanishes} for non-integer baryon number, i.e. for non-zero triality sectors.
Note that our argument does not rely on particular spatial boundary conditions. The same conclusion holds
for periodic or, for example, $C$-periodic spatial b.c., even though the latter break the $Z(3)$ symmetry. 
For convenience, we write
\be \label{eq:canonicalpartitionfunction}
  Z_{C}(T,B)  = \frac{1}{2\pi} \int_{-\pi}^{\pi} d\left(\frac{\mu_I}{T} \right)  e^{-i 3 B \frac{\mu_I}{T} }Z_{GC}(T,i \mu_I)\;.
\ee

Now, the fugacity expansion allows us to go back from the canonical partition functions to the grand
canonical one:
\be
Z_{GC}(T,\mu) = \sum_{Q=-\infty}^{+\infty} e^{Q \frac{\mu}{T}} Z_C(T,Q)\;.
\ee
This expression is identical to Eq.(\ref{eq:introduction_ZGC}), as can be seen by substituting (\ref{eq:ZC_Q})
for $Z_C(T,Q)$ above and summing over $Q$ first. However, we can remove from the sum the non-zero triality
(fractional $B$) sectors, since each of them gives a zero contribution, thus obtaining:
\be
\label{eq:ZGC_fugexp}
Z_{GC}(T,\mu) =\sum_{B=-\infty}^{\infty} e^{3 B  \frac{ \mu }{ T }} Z_C(T,B)\;.
\ee

Let us stress again that this grand canonical partition function is identical with the one given in Eq.(\ref{eq:introduction_ZGC}).
The two expressions differ by the inclusion of non-zero triality sectors, which we just showed are zero. 
However, these zero contributions are explicitly projected out in (\ref{eq:ZGC_fugexp})~\cite{Oleszczuk:yg}. 
In the following, we will come back to this and consider the effect of this projecting out on the 
$Z_3$-sensitive Polyakov loop.

\section{Polyakov Loop in the Canonical Ensemble}
\label{sec:polyakovloop_canonical}

\begin{figure}[hbt]
\begin{center}
\includegraphics[width=13.5cm]{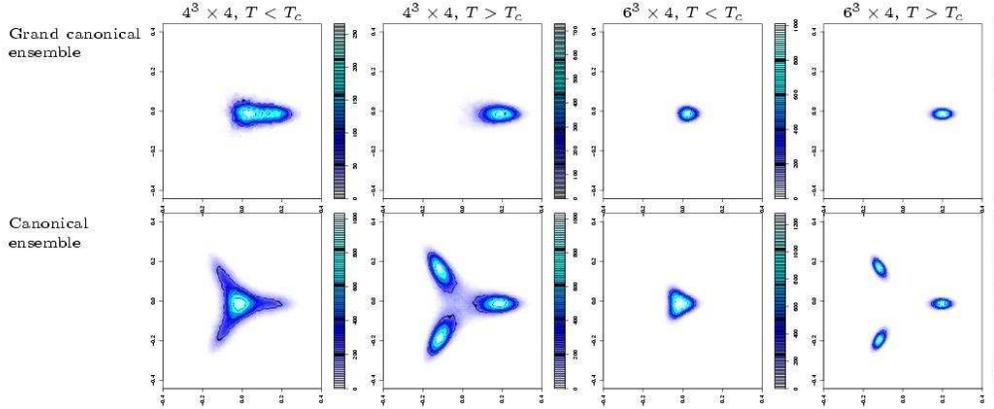}
\caption{Distribution of the complex Polyakov loop trace in the grand
canonical ({\em top}) and canonical ({\em bottom}) ensembles. {\em left}: $4^3\times
4$, {\em right}: $6^3\times 4$. In the thermodynamic limit, the
distributions agree for both ensembles, up to two additional
$Z_3$-rotations in the canonical ensemble.} \label{fig:polloop}
\end{center}
\end{figure}

The expectation value of the Polyakov loop in the canonical ensemble is zero for all temperatures and volumes.
We show this explicitly as follows. The chemical potential is introduced on the lattice
as an external imaginary gauge field
\begin{align}
 U_4(x) \to& e^{+\mu a} U_4(x)\\
 U^\dagger_4(x) \to& e^{-\mu a} U^\dagger_4(x)\;,
\end{align}
or equivalently as
\begin{align}
 U_4({\vec x},x_4=x_{4_0}) \to e^{+ N_t \mu a} U_4({\vec x},x_4=x_{4_0})\\
 U^\dagger_4({\vec x},x_4=x_{4_0}) \to e^{- N_t \mu a} U^\dagger_4({\vec x},x_4=x_{4_0})\;,
\end{align}
on a given temporal hyperplane $x_{4_0}$.
An imaginary chemical potential $i \mu_I = i \frac{2 \pi T k}{3}$ can then be absorbed in a $Z_3$ centre transformation
\be
   U_4({\vec x},x_4=x_{4_0}) \to  e^{i N_t a \frac{2 \pi T k}{3 }} U_4(x) = z(k) U_4({\vec x},x_4=x_{4_0})\;.
\ee
with $z(k) \equiv e^{i \frac{2\pi k}{3}}\; \mathbbm{1}_{3}$.
As a consequence, the two configurations $\{U, \mu_I \}$ and $\{z(k) U, \mu_I - \frac{2 \pi T k}{3}\}$
have the same value for the Dirac determinant $\det M(U; \mu_I) = \det M(z(k) U; \mu_I - \frac{2 \pi T k}{3})$,
but the Polyakov loop is centre-rotated.
We can then group the configurations of a canonical ensemble in triplets having 
$Z_3$-rotated Polyakov loop:
\be
Z_C(T,B) =\frac{1}{2\pi}\int_{-\pi}^{\pi} d\left(\frac{\mu_I}{T}\right)\; e^{-i 3B \frac{\mu_I}{T}}
    \int [DU]\;e^{-S_g[U;\beta]}\;\frac{1}{3}\sum_{k=0}^2 \det M(z(k) U_4(x_4=x_{4_0}), \mu_I)\;,
\ee
The three members of a triplet give identical contributions to $Z_C(T,B)$, 
since $\det M(z(k) U; \mu_I) = \det M(U; \mu_I+ \frac{2 \pi T k}{3})$ and $e^{-i 3B \frac{1}{T}\frac{2 \pi T k}{3}}=1$
for $B \in \mathbb{Z}$. In Fig.~\ref{fig:polloop}, we show the distribution of the Polyakov loop, where
the $Z_3$ symmetry (and hence the triplets) is clearly visible in the canonical ensemble (bottom).
In each triplet, the average of the Polyakov loops is
$Pol_i \times \left(1 + e^{-i \frac{2 \pi}{3}} + e^{i \frac{2 \pi}{3}}\right) =0$, and therefore the 
ensemble average also vanishes:
\be \label{eq:polBis0}
   \langle Pol \rangle_{Z_C(T,B)}=0
\ee
for any integer baryon number and temperature. Note again that the argument does not depend on a particular
choice of spatial boundary conditions.

\section{Polyakov Loop in the Grand Canonical Ensemble}
\label{sec:polyakovloop_grandcanonical}

In the ensemble generated by the grand canonical partition function Eq.(\ref{eq:introduction_ZGC})
\be
Z_{GC}(T,\mu) = \int [DU] e^{-S_g[U;\beta]} \det M(U;\mu)\;,
\ee
the fermion determinant explicitly breaks the $Z_3$ symmetry, so that the expectation value of the Polyakov loop
\be
   \langle Pol \rangle_{Z_{GC}(T,\mu)} \neq 0
\ee
is non-zero for any chemical potential, temperature and volume. In the following, we
show that this non-vanishing value is caused by canonical sectors 
with quark numbers which are not a multiple of three.\\

We express the grand canonical partition function via the fugacity expansion in the quark number $Q$
(we take $\mu=0$ for notational simplicity only; the argument holds for any $\mu$):
\begin{align}
Z_{GC}(T,\mu=0) &= \sum_Q Z_C(T,Q) \text{\;\;with\;\;} Z_C(T,Q) = 0 \text{\;\;if\;\;} Q \neq 0\;\text{mod}\;3 \notag  \\
& = ... + Z_C(0)+ \slashZ{Z}_C(1) + \slashZ{Z}_C(2) + Z_C(3) +
\slashZ{Z}_C(4) + ...\;,
\end{align}
where $\slashZ{Z}_C(\cdot)$ indicates $Z_C(\cdot)=0$.
The canonical partition functions can be written as $Z_C(T,Q) = \sum_i W_i(Q)$, where $i$ labels
each configuration, and $W_i(Q)$ is the corresponding Boltzmann weight.
The expectation value of the Polyakov loop is then generically given by
\begin{align}
\langle Pol \rangle_{GC} &=\frac{\sum_Q \text{\;Num}(Q)}{Z_{GC}(T,\mu=0)} \\
  &= \frac{... + \slashN{\text{\;Num}}(0) + \text{\;Num}(1) + \text{\;Num}(2) + \slashN{\text{\;Num}}(3) + \text{\;Num}(4) + ..}
{... + Z_C(0) + \slashZ{Z}_C(1) + \slashZ{Z}_C(2) + Z_C(3) +
\slashZ{Z}_C(4) + ...} \neq 0\;, \notag
\end{align}
where $\textnormal{Num}(Q) = \sum_i Pol_i W_i(Q)$, which vanishes if $Q$ is a multiple of 3 due
to Eq.(\ref{eq:polBis0}). It follows that the contributions of canonical sectors with fractional baryon number to the Polyakov loop
are unphysical, since the corresponding canonical expectation value is infinite:
\be
\langle Pol \rangle_{Z_C(T,Q \neq 0\;\text{mod}\;3)} = 
\frac{\text{\;Num}(Q \neq 0\;\text{mod}\;3)}{\slashZ{Z}_C(Q \neq 0\;\text{mod}\;3)} = 
\frac{\text{non-zero}}{0} =
\infty\;.
\ee

This argument makes it clear that the expectation value of the Polyakov loop is non-zero for all temperatures and
volumes, if we use the grand canonical partition function with respect to the \emph{quark} number, see Eq.(\ref{eq:introduction_ZGC})
and Fig.~\ref{fig:polloop}, top row.
However, if we use the grand canonical partition function with respect to the \emph{baryon} number, see Eq.(\ref{eq:ZGC_fugexp}),
the expectation value of the Polyakov loop will be exactly zero even in this equivalent grand canonical formulation.

Thus, the physical meaning of the $Z_3$-sensitive Polyakov loop expectation value is rather limited. 
It is the $Z_3$-invariant {\em correlator} $\langle Pol(0) Pol(x)^\dagger \rangle$ which is physical,
and indicates confinement or deconfinement by its $|x| \to \infty$ limit.
In the canonical ensemble, this limit is not equal to $|\langle Pol \rangle|^2$, which is identically zero:
the clustering property is not satisfied.
This is evidence of spontaneous breaking of the center 
symmetry\footnote{We are grateful to L. Yaffe for pointing this out to us.}
in the presence of fermions. The symmetry is broken spontaneously at all temperatures and densities,
rather than explicitly as in the usual grand canonical ensemble.

\section{Numerical Approach to Zero Baryon Density}
\label{sec:numerical_approach}

In order to design an algorithm that is able to measure an observable as a function of the quark, or rather
baryon number, we need to understand how the expectation value of an observable $\hat O$ can be evaluated
in the canonical ensemble. It is given by

\be\label{eq:canonicalexpvalue}
\langle \hat O \rangle_B \equiv \frac{ \frac{1}{2\pi}\int_{-\pi}^{\pi}\; d \bar \mu_I \;
e^{-i 3 B \bar \mu_I} \int [DU] \; e^{-S_g[U; \beta]} \det M(U;i\mu_I=i \bar \mu_I T)\; \hat O(U)}{Z_C(T,B)}\;.
\ee
We recognise the following numerical description. We treat $\bar \mu_I$ as a dynamical degree of freedom,
and supplement the ordinary Monte Carlo (Hybrid MC,
R-algorithm, PHMC, RHMC, $\ldots$) at fixed $\bar \mu_I$ with a noisy
Metropolis update of $\bar \mu_I \rightarrow \bar \mu_I'$ keeping the configuration
$\{U\}$ fixed. Thus, we alternate two kinds of Metropolis steps.
\begin{enumerate}
\item[(i)] Update of the links by standard Hybrid Monte Carlo~\cite{Duane:1987de}:\\
Keeping the imaginary chemical potential $\mu_I$ fixed,
we propose a new configuration $\{U'\}$, obtained by leapfrog integration of Hamilton's equations, 
as a Metropolis candidate. It is then accepted with the ordinary Metropolis probability
\be
  Prob(U \to U') = \min\left(1,e^{ - \Delta S} \right)\;,
\ee
where $\Delta S$ is the difference between the action of $\{U'\}$  and that of $\{U\}$.
\item[(ii)] Metropolis update of the imaginary chemical potential by a noisy estimator:\\
Keeping the gauge field configuration $\{U\}$ fixed,
we propose a new imaginary chemical potential $\mu_I'$, obtained from $\mu_I$ by a random step
drawn from an even distribution.
The update is based on the acceptance
\be
 \textnormal{Prob}(\mu_I \rightarrow \mu_I') = \min\left(1,\frac{e^{-i3B\bar\mu_I'}\det^{N_f}(\Dslash (\mu_I') + m)}
 {e^{-i3B\bar\mu_I}\det^{N_f}(\Dslash (\mu_I) + m)}\right)\;.
\ee

The ratio of determinants is evaluated with a stochastic estimator
(see Appendix \ref{sec:estimator}), namely
\be
\frac{\det^{N_f}(\Dslash
(\mu_I') + m)} {\det^{N_f}(\Dslash (\mu_I) + m)} = \langle e^{-
|(\Dslashup (\mu_I') + m)^{-N_f/2} (\Dslashup (\mu_I) + m)^{N_f/2}
\eta| ^2 + |\eta|^2 }   \rangle_\eta
\ee
where $\eta$ is a Gaussian complex vector. Since one Gaussian vector is
sufficient, the computational overhead is negligible.
\end{enumerate}

The algorithm above allows the sampling of any positive measure in $\mu_I$. However,
the oscillatory part $e^{-i 3 B \bar \mu_I}$ in the sampling weight causes a sign problem for non-zero baryon
number $B$. One can use $|\cos (3 B \bar \mu_I)|$ as a sampling measure and fold the remaining sign in the
observable. But such an approach breaks down for rather small $B$ already~\cite{Alford:1998sd}.
Nevertheless, at $B=0$, $e^{-i 3 B \bar \mu_I}|_{B=0}=1$ and thus, the Boltzmann weight is real and positive.
Thus, this simple algorithm suits our purpose here.
To help ergodicity, we can also perform a ${\text{``$Z_3$-move''}}$ at any time:
\begin{align}
{\mu_I \to \mu_I \pm \frac{2\pi T}{3}}&& {U_4(\vec x,x_4=x_{4_0}) \to
U_4(\vec x,x_4=x_{4_0}) e^{\mp i \frac{2\pi}{3}},\;\forall \vec x}\;,
\end{align}
where $U_4(\vec x,x_4=x_{4_0})$ are the temporal links at a given time-slice $x_{4_0}$.
Such a ``$Z_3$-move''  is always accepted, since the configuration
 $\{U, \mu_I \}$ and the one with a centre-rotated Polyakov  loop,
but shifted imaginary chemical potential, $\{U \times e^{-i \frac{2 \pi}{3}}, \mu_I + \frac{2 \pi T}{3}\}$
have the same Dirac determinant, and thus the same sampling weight,
as discussed at length in section  \ref{sec:polyakovloop_canonical}.\\

A computational detail: For $T>T_c$, the $\mu_I$-distribution is sharply peaked around $0, \pm \frac{2\pi T}{3}$.
To sample this distribution accurately in the whole interval, we apply a multicanonical algorithm in the $T>T_c$ regime
for the larger lattices ($6^3\times 4$ and $8^3 \times 4$)~\cite{Berg:1992qu}. For this, we bias the sampling of the imaginary
chemical potential by modifying the acceptance probability
\be
 \textnormal{Prob}^{\textnormal{multi}}(\mu_I \rightarrow \mu_I') =
 \min\left(1,\frac{\det^{N_f}(\Dslash (\mu_I') + m)} {\det^{N_f}(\Dslash (\mu_I) + m)}\; e^{(bias(\mu_I')-bias(\mu_I))}\right)\;,
\ee
with $bias(\mu_I)$ chosen such that the sampled histogram becomes flat for all $\mu_I$\footnote{A simple way to get an
estimate of the function $bias(\mu_I)$ is the following: One starts by sampling with no bias to produce a histogram $hist(\mu_I)$ of the
sampled $\mu_I$. One then fits $bias(\mu_I)$ to $-\log(hist(\mu_I))$ with a suitable Ansatz like $a \mu_I^2 - b \mu_I^4$, or uses a table.}.
The expectation value of an observable $\hat O$  is then given by
\be
  \langle \hat O \rangle = \frac{1}{\sum_{ \{U;\mu_I\}}e^{-bias(U;\mu_I)}}
      \sum_{\{U;\mu_I\}} \hat O(U;\mu_I) e^{-bias(U;\mu_I)}\;,
\ee
where $\{U;\mu_I\}$ labels the configurations $\{U\}$ sampled at imaginary chemical potential $\mu_I$.\\

We focus on four flavours of Kogut-Susskind fermions with degenerate mass $a m = 0.05$ and
lattices with $N_t=4$ time-slices, i.e. $\frac{m}{T}=0.2$.
With these parameters, the zero-temperature pion mass is about 350 MeV~\cite{Lombardo}.
Simulations are performed on lattices with spatial extents $4^3$, $6^3$ and $8^3$ at seven
temperatures\footnote{We relate the coupling $\beta$ to the temperature $T$
via $T=\frac{1}{a(\beta) N_t}$ and the perturbative two-loop $\beta$-function.},
ranging from $\frac{T}{T_c}=0.85$ to $1.1$, with good overlap between the ``neighbouring''
ensembles. We analyse the results using Ferrenberg-Swendsen reweighting~\cite{Ferrenberg:yz}.

\section{Free Energy Density}
\label{sec:freeenergy}

In the grand canonical ensemble, the change of the free energy density with chemical potential $\mu$
(as a dimensionless quantity) is given in terms of the partition function
\be
 \frac{\Delta F(T,\mu)}{V T^4} \equiv - \frac{1}{V T^3} \log \frac{Z_{GC}(T,\mu)}{Z_{GC}(T,0)}\;.
\ee
A standard approach~\cite{Allton:2002zi,Allton:2003vx} is to perform a Taylor expansion in $\mu$ about $\mu=0$, where
the derivatives entering the series may be expressed as complicated expectation values evaluated at $\mu=0$.
Remember that this expansion is in even powers of $\mu$, since $Z_{GC}(\mu) = Z_{GC}(-\mu)$.
In our approach, the free energy density comes for free from the $\mu_I$-histogram in the canonical simulation, and moreover,
to all orders. At low temperature, however, the histograms are quite noisy. Therefore we will, when needed, use
results from the more sophisticated method~\cite{Kratochvila:2005mk}, where we can calculate
the grand canonical partition function for an arbitrary imaginary chemical potential as a consequence
of the reweighting method that we apply.\\

\begin{figure}[htb]

\begin{center}
\includegraphics[height=5.4cm,width=4.8cm,angle=-90]{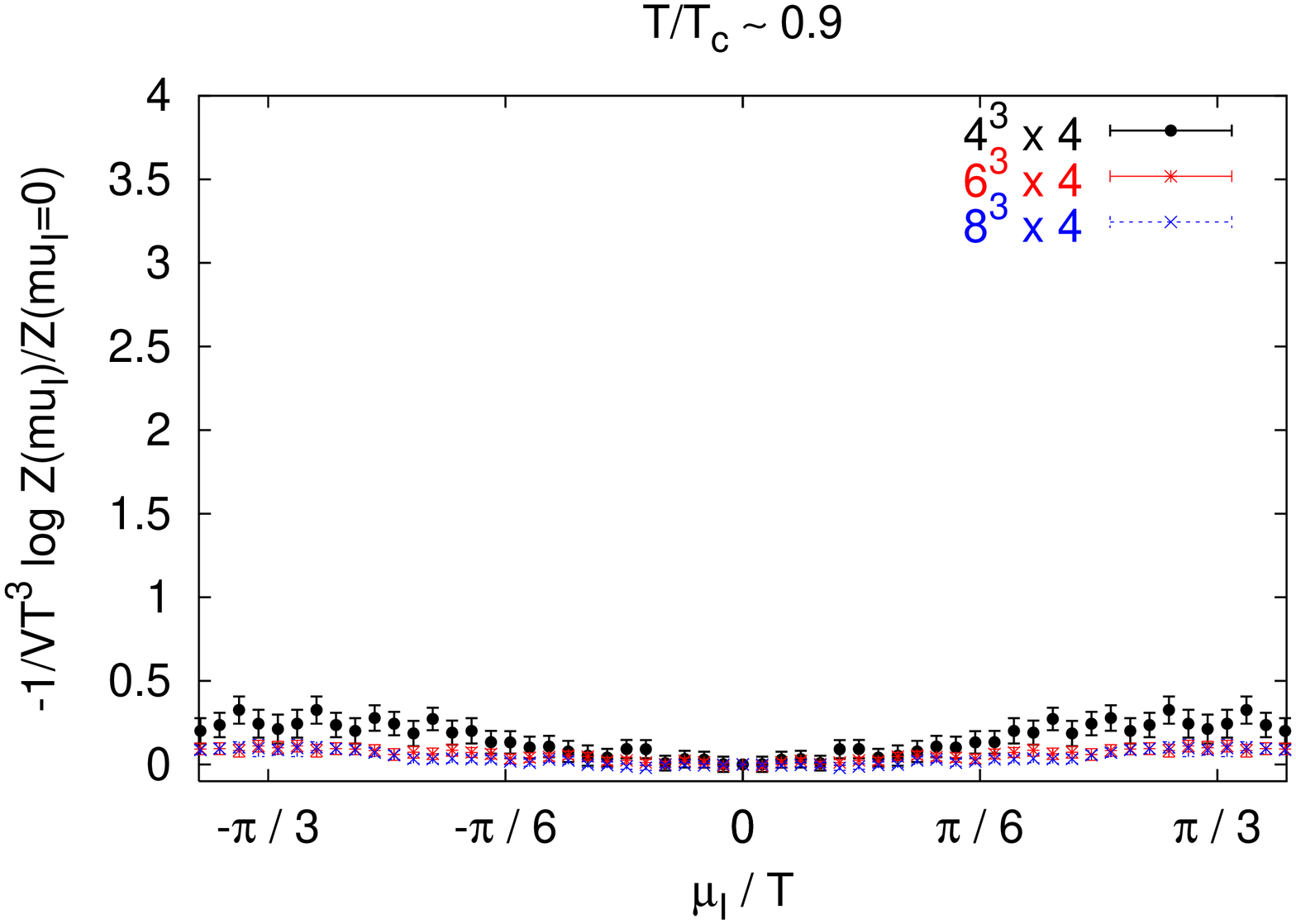}\hspace*{-0.25cm}
\includegraphics[height=5.4cm,width=4.8cm,angle=-90]{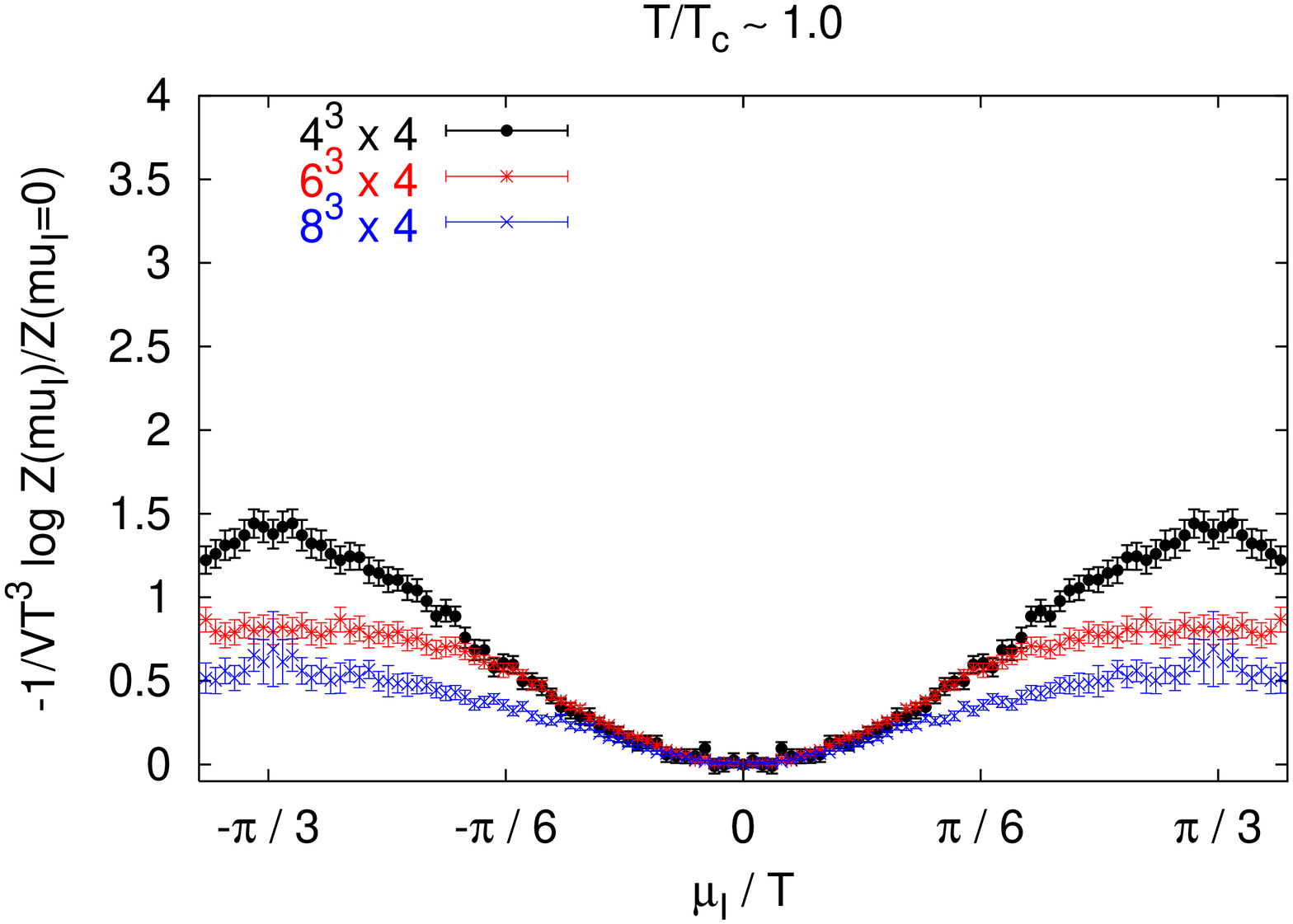}\hspace*{-0.25cm}
\includegraphics[height=5.4cm,width=4.8cm,angle=-90]{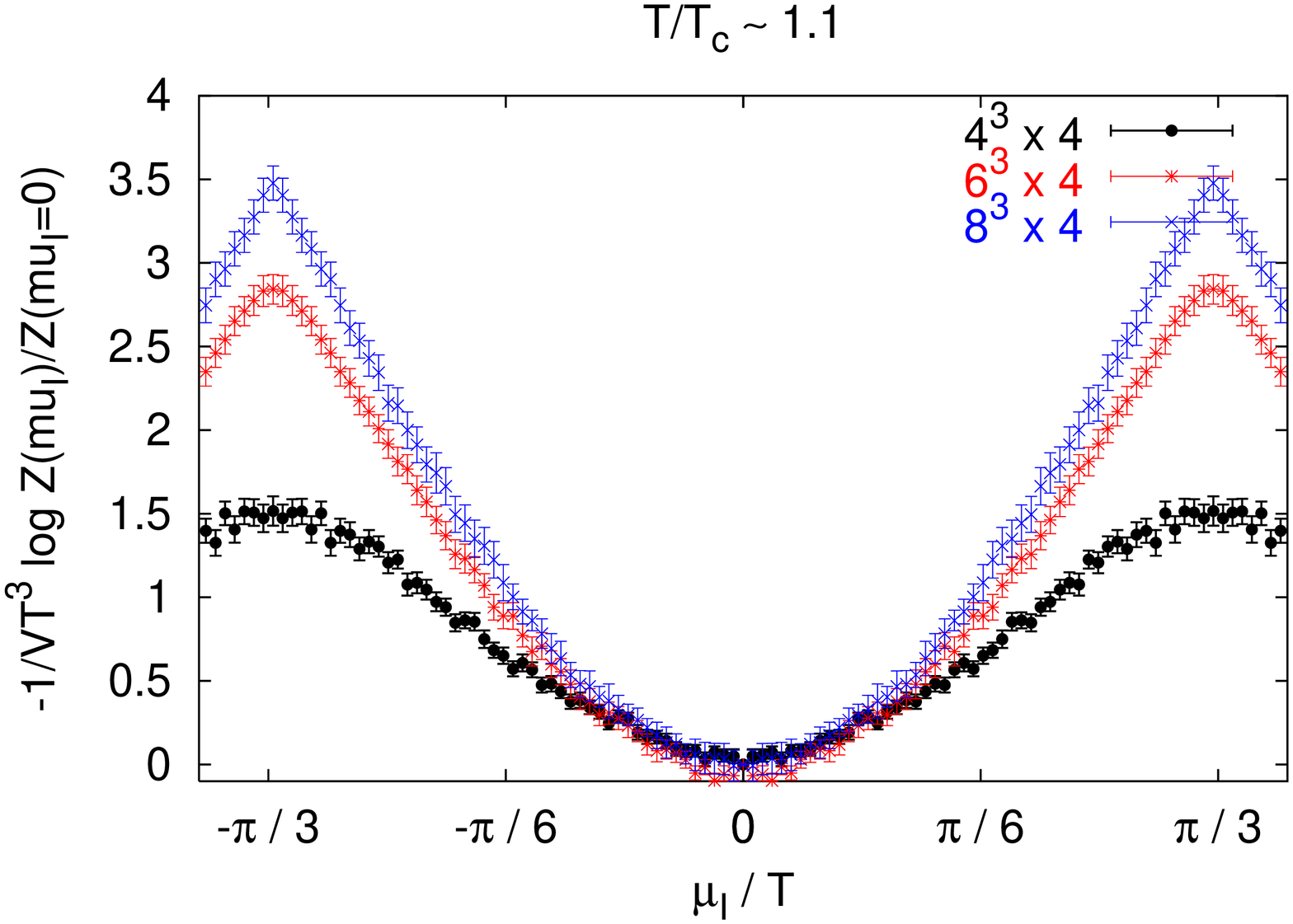}\hspace*{-0.25cm}
\caption{ $\frac{\Delta F(T,\mu_I)}{V T^4}$ as a function of $\frac{\mu_I}{T}$, at temperatures $\frac{T}{T_c}
\sim 0.9,\; 1.0,\; 1.1$ from left to right. The free energy density varies much more upon entering the
high-temperature phase, and the $Z_3$ first-order transitions become visible (right).}
\label{fig:results_chempot}
\end{center}
\end{figure}

\begin{figure}[!htb]
\centering
\includegraphics[width=5cm,angle=-90]{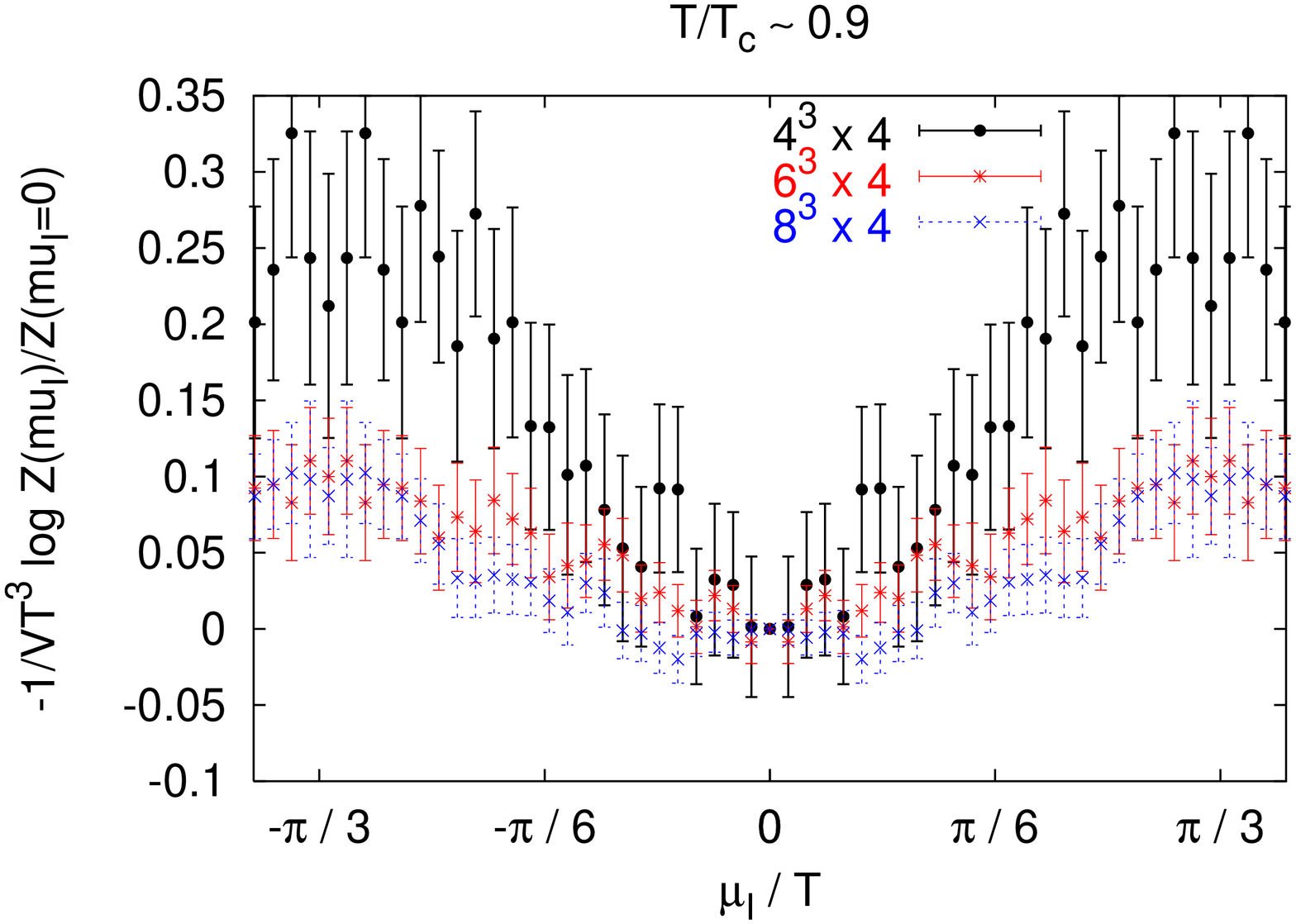}
\includegraphics[width=5cm,angle=-90]{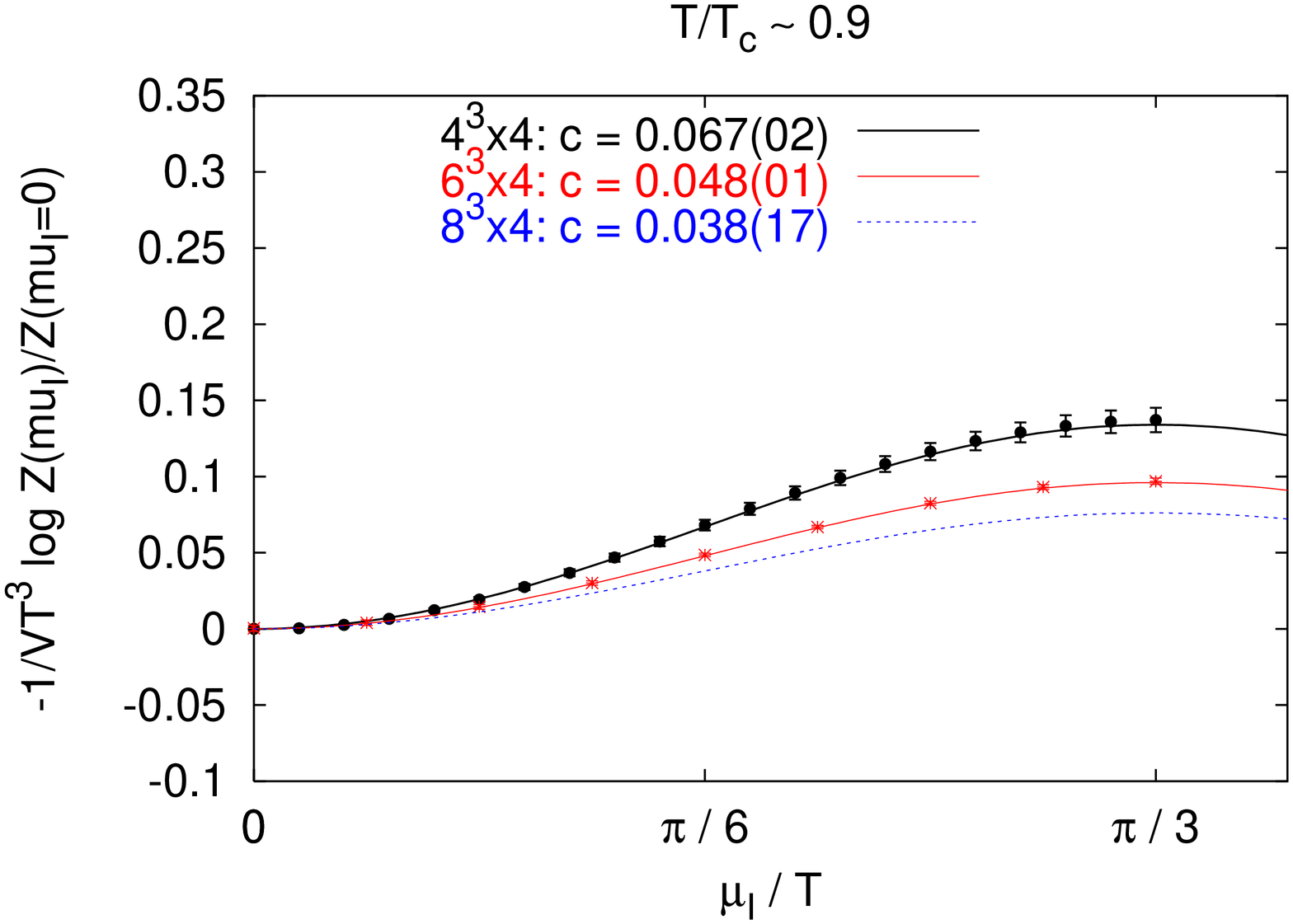}
\caption{$\frac{\Delta F(T,\mu_I)}{V T^4}$ as a function of $\frac{\mu_I}{T}$ for $\frac{T}{T_c} \sim 0.9$.
The histogram method is very noisy. We show ({\em left}) a rescaled
version of the leftmost plot in Fig.~\ref{fig:results_chempot}. We also present ({\em right})
results based on a reweighting method with variance reduction~\cite{Kratochvila:2005mk}.
The results are in agreement with the histogram method,
but allow for a more reliable description by a Fourier expansion. One Fourier coefficient suffices to describe
the data points. The reweighting method~\cite{Kratochvila:2005mk} calculation is computationally demanding and has not 
been performed yet
for the $8^3 \times 4$ lattice. We thus only draw the fit, which is based on histogram data.}\label{fig:results_chempot_495}
\end{figure}

In Fig.~\ref{fig:results_chempot} we show the free energy divided by $V T^4$
versus $\frac{\mu_I}{T}$ for $\frac{T}{T_c}  < 1$, $\frac{T}{T_c} \simeq 1$ and $\frac{T}{T_c}  > 1$.
In all cases, we observe a minimum at $\frac{\mu_I}{T}=0$. Therefore, in the thermodynamic limit, only
$\frac{\mu_I}{T}=0 \mod \frac{ 2 \pi}{3}$ will survive. This establishes numerically the expected equivalence of 
$Z_C(T,B=0)$ with $Z_{GC}(T,\mu=0)$. \\

For $\frac{T}{T_c} \sim 0.9$, no singularities develop at $\frac{\mu_I}{T} = \pm \frac{\pi}{3}$ in the thermodynamic limit,
thus indicating a crossover, as expected from the phase diagram $T$-$\mu_I$, Fig.~\ref{fig:ensemble_phasediagram}.
In Fig.~\ref{fig:results_chempot_495}, (left), we show the free energy density, determined by
the histogram method, which is very flat and noisy unfortunately. The periodicity of the free energy density
is $\frac{2\pi T}{3}$, and we exploit it by a Fourier expansion in $3k \frac{\mu_I}{T}$ using the Ansatz
\be
\frac{\Delta F(T,\mu_I)}{V T^4} = c\left(1 - \cos(3 \frac{\mu_I}{T})\right) + d \cos(6 \frac{\mu_I}{T}) + \ldots\;.
\ee
In order to improve the determination of the coefficients $c$,$d$,$\ldots$, we use results based on
the reweighting method described in Ref.~\cite{Kratochvila:2005mk}. Within errors, the free
energy density is in agreement with the histogram method, but with much smaller statistical uncertainty.
The fit is excellent already with one Fourier coefficient, with no indication for higher Fourier components,
at least on the small lattices we consider.\\

In the hadron resonance gas model (see Ref.~\cite{Allton:2005gk} for a detailed discussion),
the partition function
can be split into mesonic and baryonic contributions. Since we are interested in the dependency on a baryon
chemical potential $\mu_B=3 \mu$, it is sufficient to study the baryonic part only. In the limit $m_B \gg T,\mu_B$,
where $m_B$ corresponds to the baryonic resonance mass, the Ansatz for the free energy
density as a function of an imaginary chemical potential is
\be \label{eq:hadrongas}
\frac{F(T,\mu_I)}{V T^4}- \frac{F(T,0)}{V T^4} \equiv \frac{\Delta F(T,\mu_I)}{V T^4} = f(T) (1-\cos(\frac{3 \mu_I}{T}))\;,
\ee
where $f(T) \equiv \frac{1}{\pi^2} \sum_{i \in \mathrm{Baryons}} \left(\frac{m_i}{T} \right)^2 K_2\left(\frac{m_i}{T}\right)$.
We thus have a mean to measure the sum of resonances $f(T)$. For example in the
case of a $6^3 \times 4$ lattice, we find $f(T \sim 0.9 T_c)=0.048(1)$\footnote{We thank D.~Toublan~\cite{Toublan} for estimating
the sum of resonances for the four-flavour continuum theory. However, the result, $f(T \sim 0.9 T_c) \approx 0.2$, differs
from our determination by about a factor 4. It is unclear what is the main reason for this discrepancy, but 
the small, coarse lattice we use ($a \sim 0.3$ fm) certainly contributes an important part.}. 
Our data can be well described by this Ansatz.
This confirms our expectation that the relevant degrees of freedom in the low-temperature
phase are hadrons. The masses of these hadrons are much larger than the scale given by the temperature,
since the free energy density changes only slightly when varying the imaginary chemical potential,
thus $m_H \gg \mu_I \sim T_c \approx 160$ MeV. \\

\begin{figure}[!htb]
\centering
\includegraphics[width=5cm,angle=-90]{pictures/pressure_510_colour.eps}
\includegraphics[width=5cm,angle=-90]{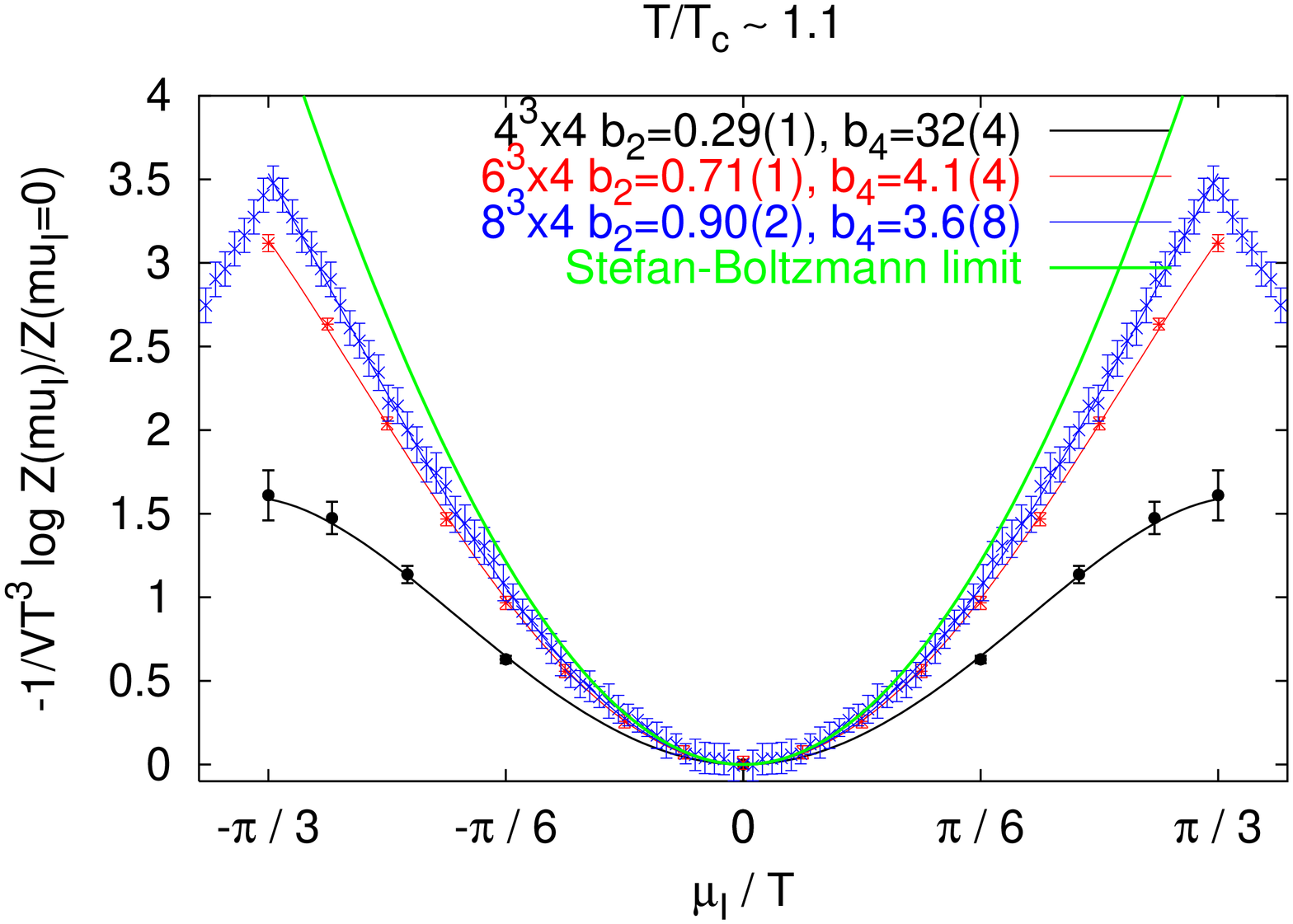}
\caption{$\frac{\Delta F(T,\mu_I)}{V T^4}$ as a function of $\frac{\mu_I}{T}$ for $\frac{T}{T_c} \sim 1.1$. 
({\em left}) The histogram method; ({\em right}) the reweighting method~\cite{Kratochvila:2005mk}, 
supplemented by the histogram results for $8^3 \times 4$. A simple modification of the free gas expression
describes all the data.
As the volume increases, the data come close to the Stefan-Boltzmann limit ($T \to \infty$)
even though $\frac{T}{T_c} \sim 1.1$ only.  }
\label{fig:results_chempot_510}
\end{figure}

For $\frac{T}{T_c}  \sim 1.1$ we expect a cusp at $\mu_I=\pm \frac{\pi T}{3}$ ($Z_3$-transitions) to develop in the
thermodynamic limit, due to the first order phase transition. Indeed, it appears clearly as the volume increases,
see Fig.~\ref{fig:results_chempot_510} for a comparison of the histogram results (left)
versus the reweighting approach\footnote{The $8^3\times 4$-data points are taken from the histogram method.} (right).\\

We can try to describe these results by a generic Taylor series in $\frac{\mu_I}{T}$ as an Ansatz, which
can be compared with a simple model at high temperature, the free gas of massless quarks.
If we perform an analytic continuation from real to imaginary chemical potential, then the free energy
density of this model is given by
\be \label{eq:fmuIansatz_simple}
  \frac{ \Delta  F(T,\mu_I)}{V T^4}= \frac{N_f}{2}\left(\frac{\mu_I}{T}\right)^2 -
        \frac{N_f}{4\pi^2}\left(\frac{\mu_I}{T}\right)^4\;.
\ee

These simple expressions are valid in the continuum theory at very high temperature, where the coupling $g(T) \approx 0$.
On the lattice we expect finite size corrections ($N_s < \infty$) as well as cut-off
corrections ($T = \frac{1}{a N_t}$). Ref.~\cite{Allton:2003vx} has calculated the free energy of free fermions 
on a lattice having
infinite spatial size ($N_s=\infty$) but finite temporal extent ($N_t=4$). Here, we also
determine the corrections for finite spatial size $N_s=4,6,8,10$ for the free massless fermion gas on the lattice.
We set the gauge fields $A_\mu(x)=0$, ie.~the gauge links to the identity, and solve for the free energy via
\begin{align}
   \frac{\Delta F^{free}_{latt}(T,\mu_I)}{V T^4} & = -\frac{\log Z^{free}(T,\mu_I)}{V T^3} =   -\frac{\log \det M^{free} (T,\mu_I)}{V T^3} \notag \\
    & \approx C_2 \frac{N_f}{2}\left(\frac{\mu_I}{T}\right)^2 - C_4 \frac{N_f}{4\pi^2}\left(\frac{\mu_I}{T}\right)^4\;.
    \label{eq:QGP_ansatz}
\end{align}
where $C_2$ and $C_4$ are fit coefficients.
Table \ref{table:corrections_freegas} summarises the results.
\begin{table}[!hbt]
\begin{center}
\begin{tabular}{|c|c|c|c|}
\hline
Lattice &  $C_2$  & $C_4$ & $[C_6]$ \\
\hline
$4^3 \times 4$  & 4.387(1) &  0.28(3) & [-] \\
$6^3 \times 4$  & 2.628(1) &  1.70(5) & [0.0081(1)]\\
$8^3 \times 4$  & 2.315(1) &  2.25(5) & [0.0046(1)]\\
$10^3 \times 4$ & 2.250(1) &  2.49(5) & [0.0030(1)]\\
\hline
$\infty^3 \times 4$ & 2.25 & 2.6 & - \\
\hline
\end{tabular}
\caption{The prediction for the free energy density based on the free massless gas model in the continuum at high temperature suffers from finite
size and cut-off effects. The correction terms $C_2$ and $C_4$ help to quantify the systematics. The functional form in Eq.(\ref{eq:QGP_ansatz})
is sufficient, since the contribution of the additional term $\left(\frac{\mu}{T}\right)^6$ is very small.} \label{table:corrections_freegas}
\end{center}
\end{table}

The coefficients $C_2$ and $C_4$ approach their infinite volume expectation rather quickly. 
For the particular quark mass $\frac{m}{T}=0.2$ which we consider, 
the difference from the massless limit is smaller than the (fitting) errors, and thus, 
results are not presented explicitly.
Note that we have an additional column $[C_6]$: we have added the term $C_6 \left(\frac{\mu}{T}\right)^6$ to the
Ansatz Eq.(\ref{eq:QGP_ansatz}).
The coefficient $C_6$ is very small and leaves $C_2$ and $C_4$ unchanged within the errors.\\

In the end, we consider the volume-dependent lattice corrections $C_2$ and $C_4$  and measure the deviation from this free
gas model by two parameters $b_2(T)$ and $b_4(T)$. The Ansatz to describe our results in Fig.~\ref{fig:results_chempot_510} 
is thus
\be \label{eq:fmuIansatz}
  \frac{ \Delta  F(T,\mu_I)}{V T^4}= b_2(T) C_2 \frac{N_f}{2}\left(\frac{\mu_I}{T}\right)^2 -
        b_4(T) C_4 \frac{N_f}{4\pi^2}\left(\frac{\mu_I}{T}\right)^4\;.
\ee

\begin{table}[!hbt]
\begin{center}
\begin{tabular}{|c|c|c|c|}
\hline
$T\sim 1.1 T_c$ &  $b_2(T)$  & $b_4(T)$  &  $b_4(T)$ (periodic) \\
\hline
$4^3 \times 4$  & 0.29(1)  &  32(4)   & 7(1)   \\
$6^3 \times 4$  & 0.71(1)  &  4.1(4)  & 0.2(7) \\
$8^3 \times 4$  & 0.90(2)  &  3.6(8)  & 1.4(4) \\
\hline
SB limit ($T \to \infty$) & 1 & 1 & 1 \\
\hline
\end{tabular}
\caption{The coefficients of the free energy density expansion for $\frac{T}{T_c} \sim 1.1$ come close to their Stefan Boltzmann
($T \to \infty$) value.
There are two values for $b_4(T)$: the first one is the result of the chi-square fit of Eq.(\ref{eq:fmuIansatz}); the
second one (``periodic'') makes use of a periodicised Ansatz, see Eq.(\ref{periodicised_ansatz}). The comparison of the two values
gives some measure of the systematic error.}
\label{table:coefficients_freegas}
\end{center}
\end{table}

We observe that the leading term approaches the Stefan Boltzmann limit
rather fast upon increasing the volume, see Fig.~\ref{fig:results_chempot_510}
and Table \ref{table:coefficients_freegas}.
This is somewhat surprising since this coincidence with the Stefan Boltzmann law
will occur only at $T \to \infty$.
Deviations at $T \sim 1.1\;T_c$ should persist even in the thermodynamic limit, reflecting the interactions
of the quarks\footnote{It has been shown already that the free energy of the gluon sector deviates
from the Stefan Boltzmann value $\frac{8\pi^2}{45}$  by about 15\%~\cite{Boyd:1996bx} even
at $\frac{T}{T_c} \sim 5$ (and more for lower temperatures). It thus would be natural to observe
deviations at finite temperature also in the quark sector.}. The reduction of $b_2(T)$ from 1 is consistent
with leading perturbative corrections~\cite{Allton:2005gk}. The value we obtain is consistent with that measured in Ref.~\cite{D'Elia:2005qu}.
The simple prediction of the free massless quark gas model works better than expected. Thus,
the relevant degrees of freedom at high temperature are very light quarks,
which is also visible in the strong dependency of the free energy density on the
imaginary chemical potential, hence $m_q \ll \mu_I \sim T_c \approx 160$ MeV.\\

The coefficient $b_4(T)$ in Table \ref{table:coefficients_freegas} suffers from  systematic
fitting errors. One source is the fitting \emph{range}: our Ansatz Eq.(\ref{eq:fmuIansatz})
does not reflect the $\frac{2\pi}{3}$-periodicity of $\frac{\mu_I}{T}$, therefore we are allowed to
fit small $\frac{\mu_I}{T}$ only. In this regime, the quartic term is subleading and hard to
quantify. An estimate of the systematic fitting error can be obtained by varying the fitting range (not explicitly
tabulated). Another source is the fitting \emph{Ansatz}: we could add the next-order
term $\left(\frac{\mu_I}{T}\right)^6$, which changes $C_4$ by a few percent, or periodicise the Ansatz by hand via
\be \label{periodicised_ansatz}
 \frac{ \Delta  F_{per}(T,\mu_I)}{V T^4} =  -\frac{1}{V T^3}  \log \frac{ Z_{per}(T,\mu_I)}{Z_{per}(T,0)}
\ee
with
\be
Z_{per}(T,\mu_I) = \sum_{k=-\infty}^\infty e^{ -VT^3
\left( b_2(T) C_2 \frac{N_f}{2}\left(\frac{\mu_I}{T} + \frac{2\pi k}{3}\right)^2 - b_4(T) C_4 \frac{N_f}{4\pi^2}\left(\frac{\mu_I}{T}+ \frac{2\pi k}{3}\right)^4 \right)}\;.
\ee
However, we must truncate (in practice, at $k=\pm 1$) the sum over all sectors to preserve convergence, 
because of the sign of the $C_4$ contribution.
In conclusion, we cannot determine $b_4(T)$ accurately. Nevertheless, it is remarkable how well the
free quark gas model describes our results. On an $8^3 \times 4$ lattice, the deviation
is only about 10\%. By using a canonical approach to simulate finite density QCD~\cite{Kratochvila:2005mk},
we can obtain more accurate results, to be presented in a follow-up publication. In particular, we can show
that $b_4(T=1.1\;T_c)=2.97(3)$.

\section{Deconfinement Transition and Finite Size Effects}
\label{sec:deconfinementtransition}

\begin{figure}[htb]

\begin{center}
\includegraphics[height=7.0cm,width=4.5cm,angle=-90]{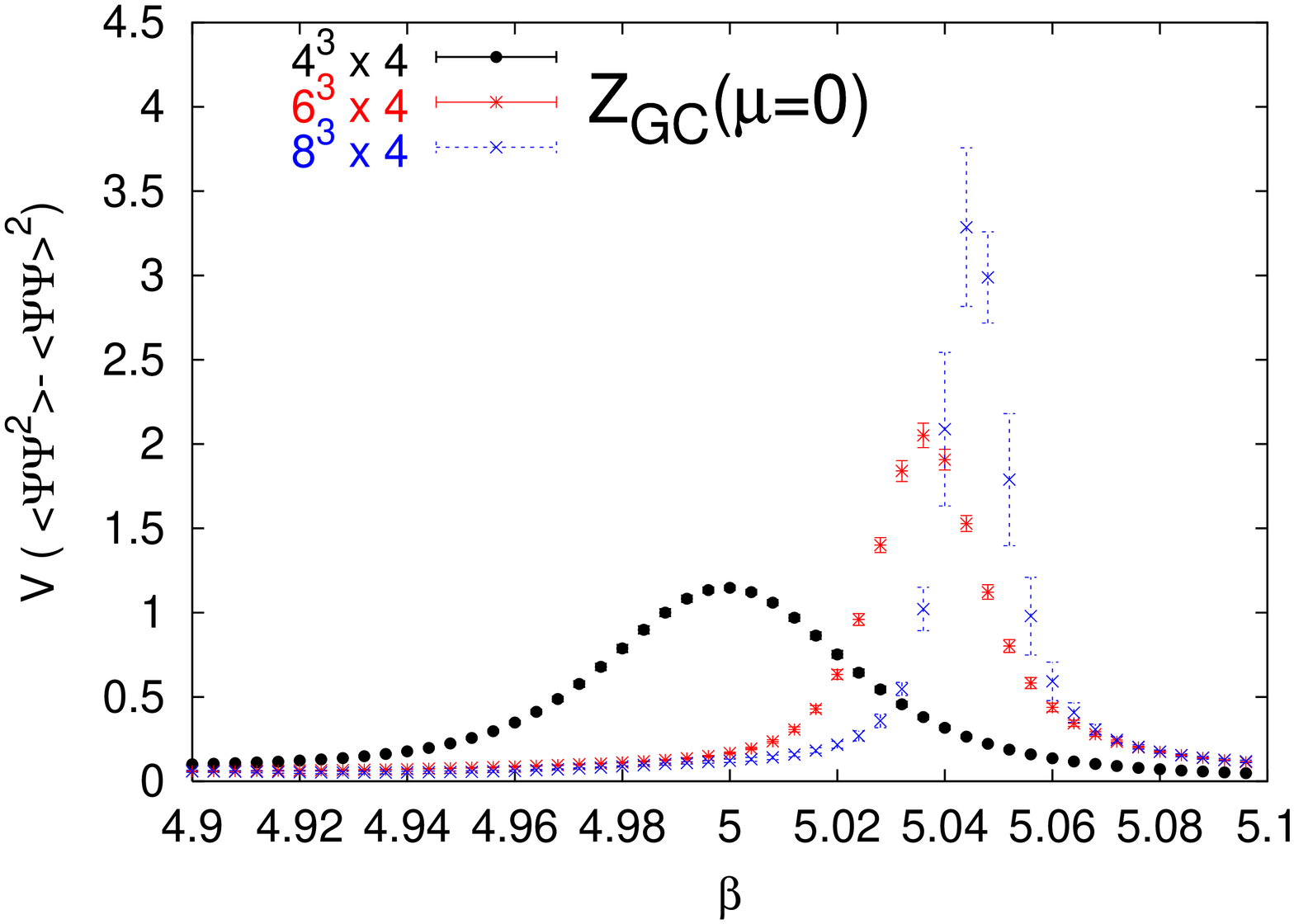}\hspace*{-0.25cm}
\includegraphics[height=7.0cm,width=4.5cm,angle=-90]{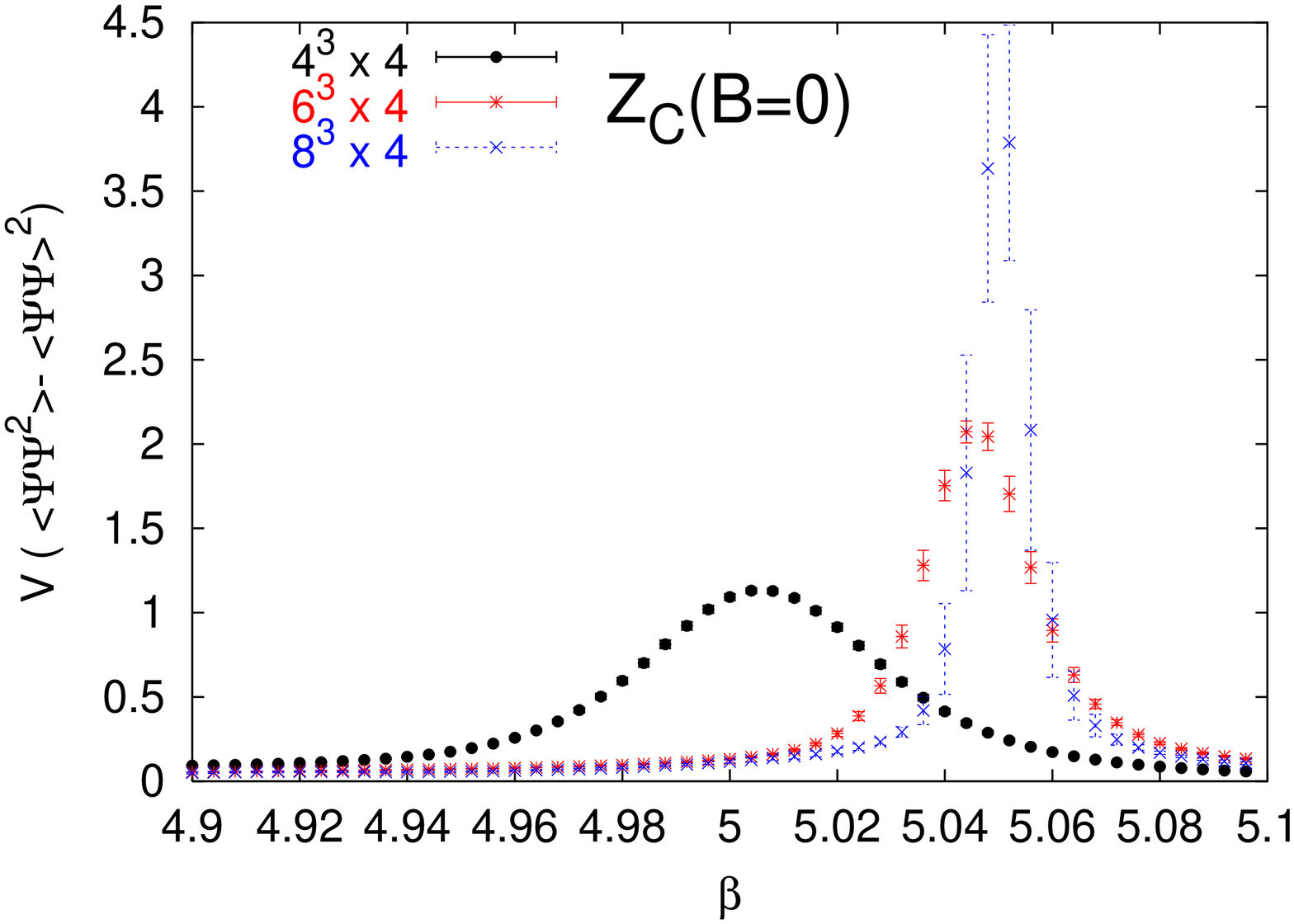}

\caption{Susceptibility of $\bar{\psi} \psi$ versus $\beta$ for
all volumes in the grand canonical and canonical ensembles. 
{\em left}: $Z_{GC}(T,\mu=0)$, {\em right}:
$Z_C(T,B=0)$. Even for the smallest, $4^4$, lattice, differences are
barely visible.} \label{fig:results_chiral}
\end{center}

\end{figure}

The phase transition is signaled by the peak in the susceptibility of the
chiral condensate  $\langle \bar{\psi} \psi \rangle$ or in the specific heat.
In Fig.~\ref{fig:results_chiral}, we show the former.

On the $4^3 \times 4$ lattice, a slight shift in the pseudo-critical $\beta_c$ can be observed between the
grand-canonical and the canonical results. It
disappears for larger volumes. We observe the same behaviour for the specific heat. The small deviation
is caused by contributions from $B \neq 0$ sectors, which are present in the $\mu=0$ ensemble. However
they are suppressed by a factor $\sim e^{-B \frac{m_B}{T}} \ll 1$, where $m_B$ is the mass of a baryon. In terms
of the baryon density $\rho$, we recognise the exponential suppression in the volume since
$e^{-B \frac{m_B}{T}} = e^{-V \rho \frac{m_B}{T}}$. Thus, we verify once more that  
the zero chemical potential ensemble is equivalent
to the zero baryon density ensemble in the thermodynamic limit.

Note that the non-zero triality sectors have zero partition function and do not contribute. They do not affect
observables studied in this section, which are insensitive to the centre symmetry. \\

\begin{figure}[htb]
\begin{center}
\psfrag{Binder cumulant minimum}{\small min $C_B(\hat O)$}
\includegraphics[height=7.0cm,width=4.5cm,angle=-90]{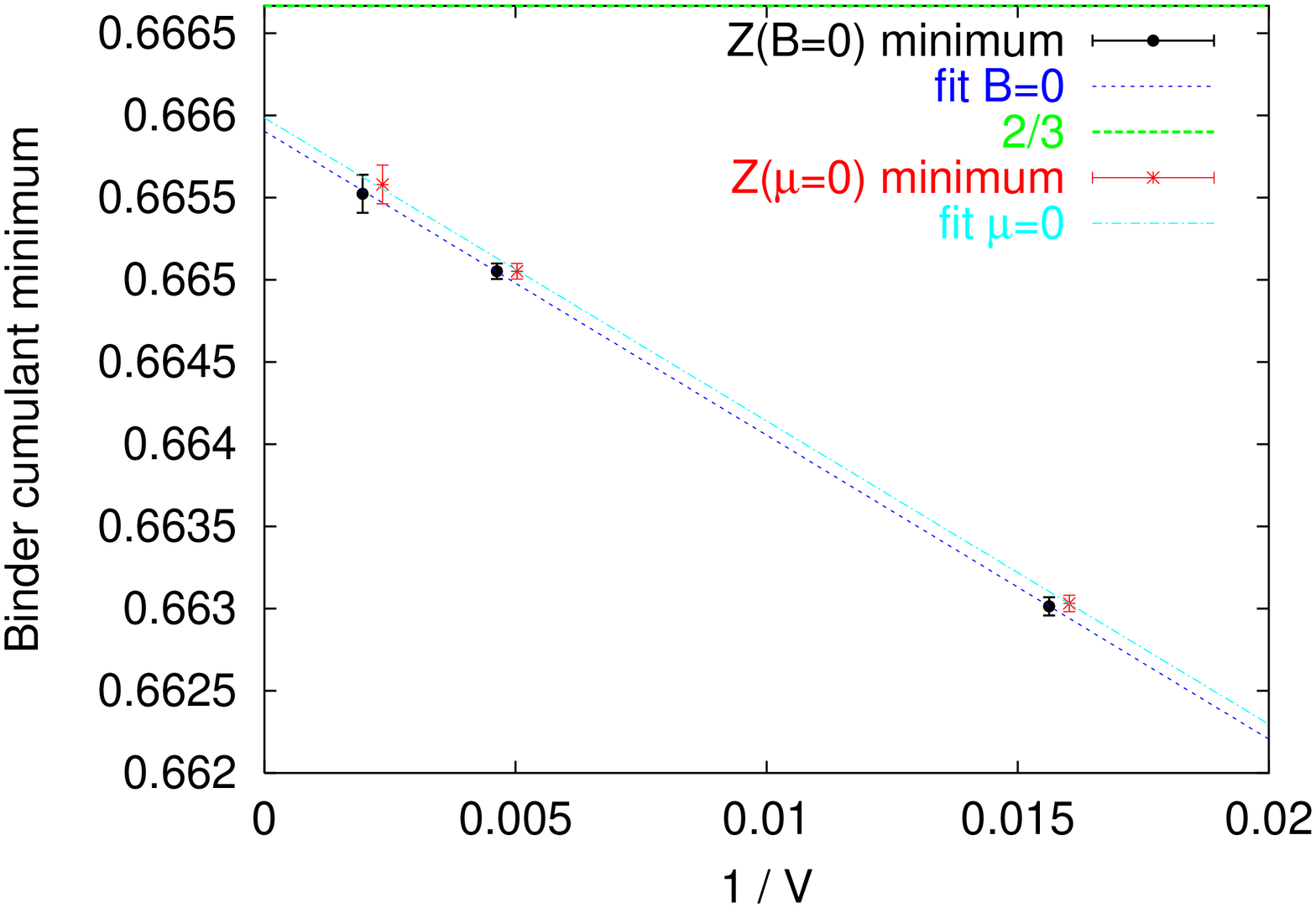}\hspace*{-0.25cm}
\includegraphics[height=7.0cm,width=4.5cm,angle=-90]{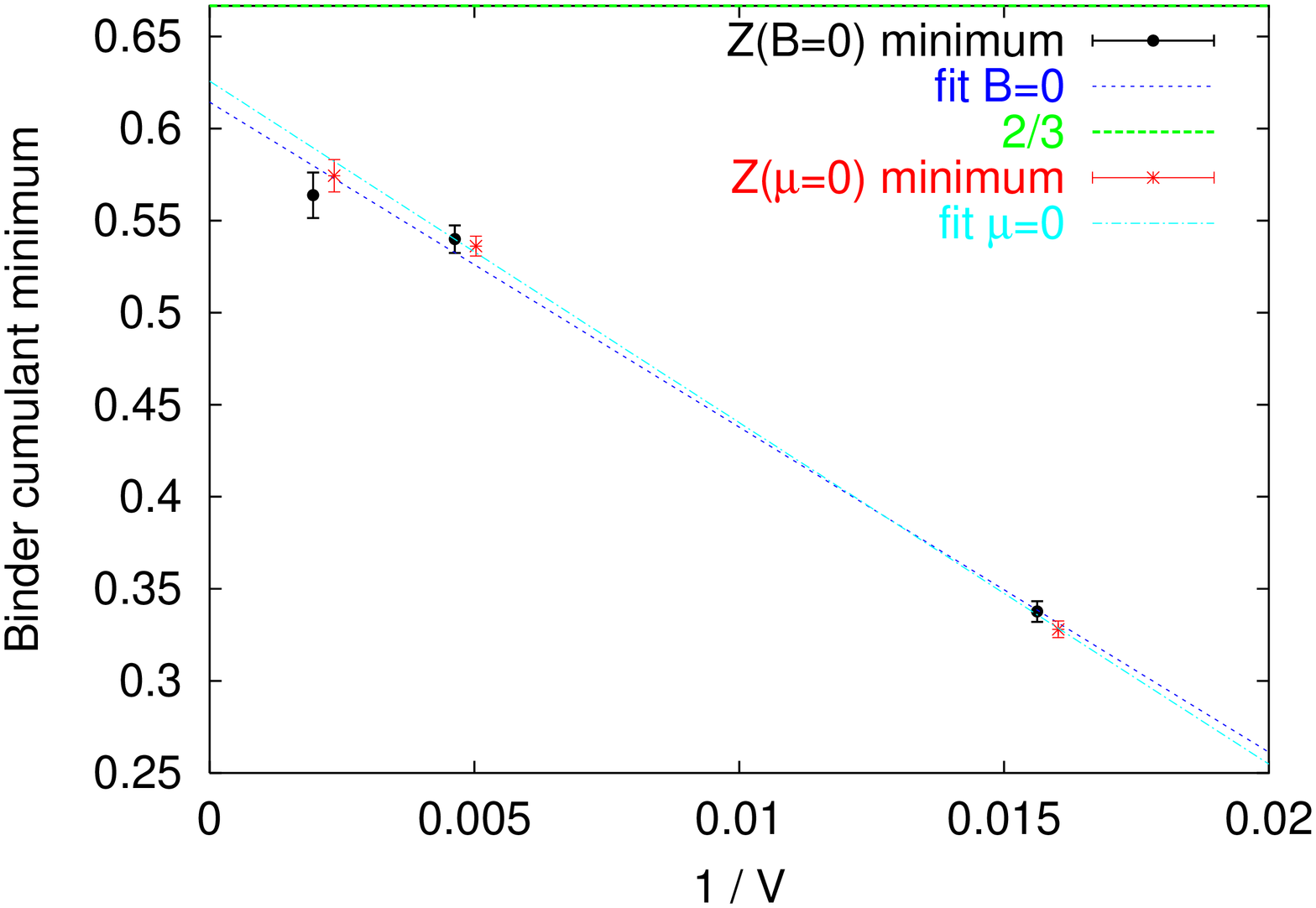}
\caption{Binder cumulant minimum versus inverse volume for both ensembles (slightly shifted in the $x$-axis to enhance
visibility). {\em left}: $\hat O=$plaquette, {\em right}: $\hat O=$chiral condensate.
The thermodynamic extrapolation does not reach $\frac{2}{3}$ (the upper left corner of the figure),
indicating a first order transition. Finite-size effects, reflected in the $1/V$ slope, are equivalent
for both ensembles.}\label{fig:results_binder}
\end{center}
\end{figure}

In quenched simulations, a $Z_3$-symmetrisation of the Polyakov loop is sometimes enforced by hand,
which is accompanied by reduced finite size effects~\cite{Martinelli:1982bm}. Therefore, we might
expect to observe a similar reduction in the canonical formalism as well compared to the grand canonical one.
To compare the finite size effects in the two ensembles, we analyse the minimum of the Binder cumulant~\cite{Binder:1981sa}
\be \label{qcdsmall_eq:binder}
C_B(\hat O) = 1 - \frac{1}{3} \frac{ \langle \hat O^4  \rangle}{\langle \hat O^2 \rangle^2}
\ee
versus the inverse volume $1/V$ (see Fig.~\ref{fig:results_binder}). For both the plaquette and
the chiral condensate, the thermodynamic (linear) extrapolation does not tend to $\frac{2}{3}$ - indicative
of a first order phase transition\footnote{In the case of a second order transition,
$\langle \hat O^4 \rangle$ is equal to $\langle \hat O^2 \rangle^2$ up to finite size corrections~\cite{Binder:1981sa}.
Thus, $C_B(\hat O) \to \frac{2}{3}$ in the thermodynamic limit. In the case of a first order transition, the double
peak structure of the distribution of the measurements causes a non-trivial value of the Binder cumulant.}, confirming
the finding in the literature~\cite{Fukugita:1986rr} for our quark masses. 
However,  for each volume, the measured cumulant values agree between the two ensembles
within statistical errors, indicating equivalent finite size effects.

\section{Conclusions}
\label{sec:conclusions}

For all densities, volumes, and (finite) temperatures, the Polyakov loop expectation value is non-zero in the grand 
canonical ensemble Eq.(\ref{eq:introduction_ZGC}), and zero in the {\em equivalent} canonical ensemble 
Eq.(\ref{eq:introduction_ZC}).
This Polyakov loop paradox has to be considered an artifact of keeping, in the grand canonical ensemble, 
sectors with quark numbers not multiple
of three. These canonical sectors, the so-called non-zero triality sectors, have zero partition function. Thus,
the non-vanishing expectation value $\langle Pol \rangle_{Z_{GC}(T,\mu)}$ in the common grand canonical formulation of
QCD at finite temperature and density, Eq.(\ref{eq:introduction_ZGC}), is irrelevant for thermodynamic properties.
The physically meaningful Polyakov loop correlator $\langle Pol(0) Pol(x)^\dagger \rangle$ behaves in the same way in both ensembles. \\

Because of quantum and thermal fluctuations, $\langle Pol(0) Pol(x)^\dagger \rangle$ tends to a non-zero
value when $|x| \to \infty$. On the other hand, $\langle Pol \rangle = 0$ in the canonical ensemble.
Thus, the clustering property is violated, which shows that the center symmetry is spontaneously
broken in the canonical ensemble, rather than explicitly broken by the fermion determinant as
in the usual grand-canonical ensemble. \\

Furthermore, an explicitly centre-symmetric grand canonical partition function $Z_{GC}(T,\mu)$, 
Eq.(\ref{eq:ZGC_fugexp}), can be
constructed from the canonical partition functions, where the contributions of non-zero triality states are
projected out. This partition function will give identical expectation values to the usual $Z_{GC}(T,\mu)$, apart
from a vanishing expectation value for the Polyakov loop.
Therefore, the non-zero triality states can be included or excluded: the thermodynamic properties of the theory
are unchanged. \\

We have shown this explicitly by comparing the grand canonical ensemble at $\mu=0$ with the canonical ensemble $B=0$.
Numerically, we have established the equivalence of $Z_{GC}(\mu=0)$ and $Z_C(B=0)$ in the thermodynamic limit by measuring
the free energy density as a function of $\mu_I$, using the histogram of the imaginary chemical potential distribution\footnote{Remember that our numerical
approach treats $\mu_I$ as a dynamical degree of freedom.}.
For all temperatures, we observe a minimum at $\frac{\mu_I}{T}=0$. At low temperature,
the free energy density of the confined phase can be rather well described by the hadron resonance gas, see Eq.(\ref{eq:hadrongas}).
We thus have a simple way to determine the sum of resonances $f(T)$. At high temperature, a slightly modified free gas
Ansatz, see Eq.(\ref{eq:fmuIansatz}), allows to account for all data points in the quark-gluon plasma phase.
We determine the finite-size and cut-off correction terms $C_2$ and $C_4$ by calculating
the free energy of the free fermion gas on the lattice and find agreement with the literature for $V \to \infty$.
By construction, the interaction coefficients $b_2(T)$ and $b_4(T)$ tend to 1 for high enough temperatures,
reproducing the Stefan Boltzmann law.
Just above $T_c$, the deviation from 1 in the leading coefficient
is about 30\% on a $6^3 \times 4$ lattice; on an $8^3 \times 4$ lattice, this deviation is about 10\% only.
This near-agreement with a non-interacting gas is unexpected at such comparatively low temperatures.

The approach to the thermodynamic limit is very similar in the canonical ($B=0$) and grand canonical ($\mu=0$) ensembles.
The susceptibility of the chiral condensate, or the specific heat, indicate the same pseudo-critical temperature
already on small volumes. A small shift, caused by contributions from non-zero baryon sectors, is visible only on 
the $4^3 \times 4$ lattice. \\

The zero-density canonical formulation requires a centre-symmetric simulation of QCD, which can be achieved very simply with
negligible computer overhead, by adding to the standard algorithm a single degree of freedom $\mu_I$ updated by Metropolis. \\

We hope to have fully clarified the (un)importance of non-zero triality states, and thus, to have put to rest
long-standing speculations. Further connections between the grand-canonical and the canonical formalisms
in the context of non-zero chemical potential/density will be the subject of a forthcoming paper.

\section{Acknowledgements}
\label{sec:acknowledgements}

We are grateful to O.~Jahn, K.~Kajantie, F.~Karsch, M.~P.~Lombardo, O.~Philipsen,
K.~Rummukainen, B.~Svetitsky, T.~Takaishi, D.~Toublan and L.G.~Yaffe for fruitful discussions
and advice.

\appendix
\section*{Appendix: Stochastic Estimator}
\label{sec:estimator}

A ratio of determinants can be estimated using a single Gaussian complex vector:
\begin{align}
\frac{\det^{N_f}(\Dslash (\mu_I') + m)} {\det^{N_f}(\Dslash
(\mu_I) + m)} &= \frac{\det^{N_f}M(\mu_I')} {\det^{N_f}M(\mu_I)}
= \frac{\int d\phi^\dagger d\phi e^{-\phi^\dagger \frac{1}{M^{N_f}(\mu_I')} \phi  }}{\int d\phi^\dagger d\phi e^{-\phi^\dagger \frac{1}{M^{N_f}(\mu_I)} \phi}} \\
&= \frac{  \int d\eta^\dagger d\eta\; |J(\phi,\eta,\mu_I)| \;
e^{-\eta^\dagger M^{N_f/2}(\mu_I) M^{-N_f/2}(\mu_I')
M^{-N_f/2}(\mu_I') M^{N_f/2}(\mu_I) \eta  }} {\int d\eta^\dagger
d\eta\; |J(\phi,\eta,\mu_I)| \; e^{-\eta^\dagger \eta}}
\end{align}
where we have substituted $\phi = M^{N_f/2}(\mu_I)\eta$.  Note,
that in the above notation, the Jacobian $|J(\phi,\eta,\mu_I)|$ is
$\det M^{N_f}(\mu_I)$, which is independent of $\eta$ and cancels out in the ratio:
\begin{align}
\frac{\det^{N_f}(\Dslash (\mu_I') + m)} {\det^{N_f}(\Dslash
(\mu_I) + m)} &= \frac{ \int d\eta^\dagger d\eta\;
e^{-|M^{-N_f/2}(\mu_I') M^{N_f/2}(\mu_I) \eta|^2}e^{-|\eta|^2+|\eta|^2} }{\int d\eta^\dagger d\eta\;e^{-|\eta|^2}  } \\
&= \langle e^{-|M^{-N_f/2}(\mu_I') M^{N_f/2}(\mu_I)
\eta|^2+|\eta|^2} \rangle_{\eta}\;.
\end{align}
$\langle \cdot \rangle_{\eta}$ tells us that $\eta$ has to be
sampled with the distribution $\int d\eta^\dagger d\eta\
e^{-|\eta|^2} $.

\end{document}